# The physical origin of hydrophobic effects

Qiang Sun[*]

*Key Laboratory of Orogenic Belts and Crustal Evolution, The School of Earth and Planetary Sciences, Peking University, Beijing 100871, China*

**Abtract:**

The strength of hydrogen bonding in water is stronger than that of van der waals interaction, therefore water may play an important role in the process of hydrophobic effects. When a hydrophobic solute is dissolved into water, an interface appears between the solute and water. To understand the mechanism of hydrophobic effects, it is necessary to study the structure of water and solute/water interface. In this study, based on the structural studies on water and air/water interface, the hydration free energy is derived, and utilized to investigate the physical origin of hydrophobic effects. According to the discussion on hydration free energy, with increasing solute size, it can be divided into the initial and hydrophobic solvation processes, respectively. In the initial solvation process, hydration free energy is dominated by the hydrogen bonding in interfacial water (topmost water layer at the solute/water interface). However, in the hydrophobic solvation process, the hydration free energy is related to hydrogen bondings of both bulk water and interfacial water. Additionally, various dissolved behaviors of solutes can be expected for different solvation processes. From this, hydrophobic effects can be ascribed to the competition

---

[*] Corresponding Author

E-mail: QiangSun@pku.edu.cn

between the hydrogen bondings in bulk water and those in interfacial water.

**Key words:**

Water, Interface, Hydrogen bonding, Hydrophobic effects, Hydration free energy

**1. Introduction**

The hydrophobic effects, describing the tendency of non–polar molecules or molecular surfaces to aggregate in an aqueous solution, are involved in many important chemical and biological processes including receptor–ligand interactions, protein folding and molecular assembly, as well as interactions in lipid membranes. To explain fundamental biophysics and biochemistry [1,2] as well as to engineer new materials [3,4], many theoretical and experimental works have been carried out to investigate the mechanism of hydrophobic effects [5-35]. However, a deep and quantitative understanding of the origin and nature of the interactions still remains elusive.

Historically, the concept of hydrophobicity arose in the context of the low solubility of non–polar solutes in water. The classical mechanism for hydrophobicity, proposed by Frank and Evans [5], and advanced by Kauzmann [6], and many others, predicts that, when solute being smaller than a nanometer (or concentrations below the onset of solute aggregation), water immediately surrounding the hydrophobic group is more "ordered" than bulk water. This is called as "iceberg" formation (or clathrate structure) around the hydrophobic hydrocarbon [5]. Another explanation is based on the application of scaled–particle theory (SPT) [8] to study the hydrophobic effects. In fact, SPT was originally developed for deriving the equation of state for hard–sphere fluids. Similar concepts can be used to predict the free energy of cavity formation in any bulk systems. According to SPT model [11], the process of forming an empty spherical cavity in water is equivalent to that of inserting a hard sphere into water. In particular, a recent theory by Lum, Chandler, and Weeks [12] highlight the different physical mechanisms of solvation of small and large hydrophobic solutes in water, which arises from the different manner in which they affect the structure of water. Based on their theory of hydrophobic solvation (LCW) [12], they predict that the crossover between small and large regime occurs on nanometer length scale. This suggests that the crossover is due to the change in the physical mechanism, from one entropy–dominated to another enthalpy–dominated [9]. Thermodynamically, the overall hydration free energy changes from growing linearly with solvated volume to growing linearly with solvated surface area [16].

Experimentally, hydrophobic effects can be investigated through the direct measurement of forces between hydrophobic surfaces, such as surface forces apparatus (SFA) and the atomic force microscope (AFM). Since the first direct force measurements between two hydrophobic surfaces in 1982 using the surface forces apparatus (SFA) [23], there have been many attempts to quantify the distance dependence of the attractive hydrophobic force [23-35]. The original experimental study by Israelachvili and Pashley concluded that the hydrophobic attraction was longer–ranged, stronger than van der waals interactions, and decayed approximately exponentially with a decay length of about 1 nm [23]. Subsequent studies provided wildly varying accounts of the range and magnitude of the hydrophobic attraction, with some work reporting an effective range of up to several micrometers [24].

In principle, the free energy has two primary components, $\Delta G=\Delta H-T\Delta S$, where $\Delta H$ and $\Delta S$ are the enthalpic and entropic changes incurred during solvation. The enthalpic part is a measure of the average potential energy of interactions between molecules, and the entropic part is a measure of the order or intermolecular correlations [7]. When hydrophobic solutes are dissolved into water, the thermodynamic functions may contain solute–solute, solute–solvent and solvent–solvent interaction energies, respectively. However, the strength of hydrogen bonding in water is stronger than that of van der waals interaction, water should play an important role in the process of hydrophobicity.

Hydrophobic interactions depend on temperature, pressure, solute size and shape, type, and concentration of additives as well as proximity to interfaces. However, the tendency for hydrophobic particles to cluster in water is one of the most fascinating aspects, which is readily understood in terms of the dependence of hydrophobic hydration free energy on solute size [16]. In this study, according to the structural studies on water and air/water interface, the hydration free energy is derived, and utilized to investigate the dissolved behaviors of hydrophobic solutes.

**2. Hydration free energy**

For a single $H_2O$ molecule, the vibrational normal modes are $2A_1$ (including asymmetric stretching vibration $v_1$ at 3657.05 cm$^{-1}$ and a bending vibration $v_2$ near 1595 cm$^{-1}$) + $B_1$ (anti–symmetric stretching vibration $v_3$ at 3755.97 cm$^{-1}$) [36], they are all Raman active. When a hydrogen bond forms between two water molecules, electron redistribution occurs. This increases the O–H bond lengths, while causing a 20–fold greater reduction in the H···O and O···O distances [37], in addition to causing a red shift of the hydrogen–bonded OH stretch frequency, the magnitude of which increases with cluster size [38]. The OH vibrations are sensitive to hydrogen bondings, and widely applied to investigate the structure of liquid water.

In recent years, many studies have been conducted on water molecular clusters. From the dependence of OH vibrations on water clusters $(H_2O)_n$ [39,40], when three–dimensional hydrogen bondings appear (n ⩾ 6), the OH stretching vibrations are mainly dependent on the hydrogen–bonded networks in the first shell (local hydrogen–bonded networks) of a water molecule, and the effects of hydrogen bonding beyond the first shell on OH vibrations are weak.

In this study, local hydrogen bonding refers to the interactions between a water molecule with neighboring molecules or with the hydrogen–bonded networks in the first coordination shell of the molecule.

When three–dimensional hydrogen bonding occurs, OH vibration is closely related to local hydrogen bonding of a water molecule. Therefore, it is reasonable to assign different OH stretching frequencies to OH vibrations engaged into various local hydrogen bondings. For a water molecule, the local hydrogen–bonded network can be differentiated by whether the molecule forms hydrogen bonds as a proton donor (D), proton acceptor (A), or a combination of both with neighboring molecules. According to our recent studies [39-41], at ambient conditions, the local hydrogen bonding motifs for a water molecule can be classified as DDAA (double donor–double acceptor), DDA (double donor–single acceptor), DAA (single donor–double acceptor), and DA (single donor–single acceptor). At 293 K and 0.1 MPa, the Raman OH stretching band of water can be deconvoluted into five sub–bands, located at 3045, 3225, 3435, 3575, 3635 cm$^{-1}$, and can be assigned to the DAA–OH, DDAA–OH, DA–OH, DDA–OH, and free OH symmetric stretching vibrations, respectively (Figure 1(a)).

Based on the above, for ambient water, the OH vibration is mainly dependent on the local hydrogen bonding of a water molecule. In fact, this can be confirmed by the pressure effects on water structure. From the Raman spectra of water up to 400 MPa at 293 K [41], the Raman OH stretching bands slightly move to lower wavenumber, but high pressure does not obviously change the profile of the Raman OH stretching band. Based on the explanation on Raman OH stretching bands, this indicates that high pressure has no obvious effects on the first shell of a water molecule. These results are in agreement with other studies [42,43].

For ambient water, a water molecule interacts with neighboring water molecules (in the first shell) through various local hydrogen–bonded networks. The hydrogen bonding structure is influenced by temperature, pressure, dissolved salt, and confined environments, which will rearrange to oppose the changes of external conditions. At 293 K and 0.1 MPa, the populations in DDAA, DDA, DAA, DA, and non–hydrogen bonding are 33.20 %, 5.66 %, 2.10 %, 57.36 %, and 1.68 %, respectively (Figure 1(a)). The average number of hydrogen bonds can approximately be determined to be 4·DDAA + 3·DDA + 3·DAA + 2·DA = 2.71.

From the explanation on Raman OH stretching bands of water (Figure 1(a)), the 3225 cm$^{-1}$ band is ascribed to OH vibration engaged in tetrahedral (DDAA) hydrogen bonding or an "unbroken hydrogen bond", and the 3635 cm$^{-1}$ band is assigned to the free OH stretching vibration. For the Raman OH stretching vibrations of water from 278 K to 328 K (Figure 1(b)), based on the van't Hoff equation, a plot of $\ln(I_{\text{free-OH}}/I_{\text{DDAA-OH}})$ versus $1/T$ will yield a straight line with a slope that is proportional to the energy difference between the hydrogen bonds (Figure 1(c)),

$$\ln \frac{I_{\text{free-OH}}}{I_{\text{DDAA-OH}}} = -\frac{\Delta H}{RT} + \frac{\Delta S}{R} \qquad (1)$$

From this, the $\Delta H$ and $\Delta S$ from DDAA (tetrahedral) hydrogen bonding to free water can be determined to be 11.35 kJ/mol and 29.66 J/mol, respectively.

As a hydrophobic solute is dissolved into water, an interface appears between the solute and water. Therefore, it is necessary to investigate the effects of solute/water interface on water structure. The OH vibrational frequency is mainly dependent on the local hydrogen bonding of a water molecule, therefore the solute mainly affects the structure of topmost water layer at the interface (interfacial water). Therefore, it is necessary to investigate the structure of interfacial water (topmost water layer at the interface).

In principle, the vibrational sum frequency generation (SFG) spectroscopy is an interface selective technique, and applied to investigate the water structure at the interface [44-51]. Recently, phase–sensitive sum–frequency generation (PS–SFG) spectroscopy is also developed by Shen et al [52-54]. to study the air/water interface. From this method, Im$\chi^{(2)}$ can directly be obtained from experimental measurements. According to our recent study [51] on the SFG intensity of air/water interface, in combination with PS–SFG studies [52-54], it can be derived that no DDAA (tetrahedral) hydrogen bonding can be found in interfacial water, and obvious structural difference can be expected across the interface.

From the above, the dissolved solute mainly affects the hydrogen bonding in interfacial water (the topmost water layer) (Figure 2). Additionally, in reference with bulk water, the loss of DDAA hydrogen bonding regarding to interfacial water layer may be related to the interfacial formation. After the ratio of interfacial water layer to volume is determined, the loss of DDAA hydrogen bonding can be determined. From this, the Gibbs energy of interfacial water can be determined as follows,

$$\Delta G_{Solute/water\,interface} = \Delta G_{DDAA} \cdot R_{Interfacial\,water/volume} \cdot n_{HB} \qquad (2)$$

where $R_{Interfacial\,water/volume}$ means the molecular number ratio of interfacial water layer to volume, and $n_{HB}$ is the average number of DDAA hydrogen bonding per molecule. For DDAA hydrogen bonding, $n_{HB}=4/2=2$. The $\Delta G_{DDAA}$ is the Gibbs free energy of DDAA (tetrahedral) hydrogen bonding of water.

When a hydrophobic solute is dissolved into water, this leads to the appearance of interface between water and solute (Figure 2). Thermodynamically, this is equivalent to form the solute/water interface, and the water molecules surrounded by the interface are expelled into bulk water. Of course, this equals to the water to be confined by the corresponding solute/water interface. Therefore, it is reasonable to ignore the work used to exclude the water molecules surrounded by the interface into bulk water. In other words, the process of inserting a hard sphere into water is equivalent to that of forming an empty spherical cavity in water. After the hydrophobic solute is regarded as an ideal sphere, the hydration free energy can be determined as,

$$\Delta G_{Hydration} = \Delta G_{Water} + \Delta G_{Solute/water\,interface} = \Delta G_{Water} + \frac{8 \cdot \Delta G_{DDAA} \cdot r_{H2O}}{R} \qquad (3)$$

where $\Delta G_{Water}$ is the Gibbs free energy of pure water, $\Delta G_{Solute/water\,interface}$ means the Gibbs free energy of interfacial water. Regarding to a sphere, the molecular number ratio of interfacial water layer to volume is calculated to be $4 \cdot r_{H2O}/R$, where $r_{H2O}$ is the average radius of an $H_2O$ molecule, and R is the radius of solute.

At 293 K and 0.1 MPa, the Gibbs free energy of water ($\Delta G_{Water}$) is -1500 cal/mol [55]. Additionally, the average volume per water molecule in ambient water is $3 \times 10^{-29}$ m$^3$, if treated as a sphere, which gives a diameter of 3.86 Å, and $r_{H2O}$ is 1.93 Å. The Gibbs free energy of DDAA hydrogen bonding ($\Delta G_{DDAA}$) is calculated to be -2.66 kJ/mol. Based on the equation (3), the hydration free energy can be determined (Figure 3).

**3. Hydrophobic effects**

In this study, hydration free energy is derived, which can be applied to investigate the mechanism of hydrophobic effects. From figure 3, when a hydrophobic solute is dissolved into water, this leads to the increase of hydration free energy. Because hydration free energy is the sum of Gibbs energy of water and interfacial water, with increasing the solute size, this can be divided into the initial and hydrophobic solvation processes, respectively. The structural transition between them can be expected to take place when $\Delta G_{Water}$ equals to $\Delta G_{Solute/water\ interface}$,

$$\Delta G_{Water} = \Delta G_{Solute/water\ interface} \quad (R = R_c) \quad (4)$$

where the corresponding radius of hydrophobic solute at structural transition is termed as critical radius (Rc). At 293 K and 0.1 MPa, the critical radius of dissolved solute can be determined to be 6.5 Å (Figure 3).

In the initial solvation process, the hydration energy is obviously dependent on the size of solute. However, hydration free energy is slightly independent on the size of solute in the hydrophobic process (Figure 3). From equation (3), the Gibbs energy of water ($\Delta G_{Water}$) is constant at specific temperature and pressure, therefore the different dependence of hydration free energy can be ascribed to whether hydration free energy is largely dependent on the Gibbs energy of solute/water interface ($\Delta G_{Solute/water\ interface}$) or not. In addition, the hydrogen bondings in water are different from those of interfacial water. Therefore, the dependence of hydration free energy should be closely related to the structural changes of hydrogen bondings in the process of hydrophobic effects.

In principle, the lower hydration free energy, the more thermodynamically stable. From equation (3), the hydration free energy is closely related to the size of solute (1/R), which is proportional to the ratio of surface area to volume of dissolved solutes. Therefore, hydration free energy can be utilized to investigate the dissolved behaviors of hydrophobic solutes. From this, different dissolved behaviors of solutes can be expected in the initial and hydrophobic solvation processes, respectively (Figure 4).

**3.1 Initial solvation process**

In reference to pure water, as the hydrophobic solute is dissolved into water, this leads to the increase of hydration free energy. When $\Delta G_{Solute/water\ interface}$ is less than $\Delta G_{Water}$ (both of them are negative), this is termed as the initial solvation process,

$$\Delta G_{Solute/water\ interface} < \Delta G_{Water}\ (R < R_c) \qquad (5)$$

With increasing solute size, this leads to the rapid increase of hydration free energy. This means that even small solute also affects the hydrogen bondings in water. This is in agreement with recent Kim et al. [56] study.

In initial solvation process, the hydration free energy is dominated by $\Delta G_{Solute/water\ interface}$, which is proportional to the ratio of surface area to volume of dissolved solutes (1/R). To maximize $|\Delta G_{Solute/water\ interface}|$, this may be fulfilled by the maximizing the ratio of surface area to volume of hydrophobic solutes. It can be expected that the dissolved solutes may be surrounded by interfacial water, and tend to be dispersed in water (Figure 4). Therefore, the hydration free energy may be proportional to the number of solute (or volume).

At ambient condition, DDAA and DA are primary hydrogen bonindgs in bulk water [40]. However, DA is the primary structural motif in interfacial water [51]. Therefore, in the initial solvation process, the hydration free energy is mainly dominated by DA hydrogen bondings in the interfacial water. In reference to DDAA hydrogen bonding, DA structural motif owns lower hydrogen bonding energy and higher entropy [40]. Thermodynamically, in the initial solvation process, the drive force should originate from the increase of entropy arising from interfacial water.

**3.2 Hydrophobic solvation process**

In this study, when $\Delta G_{Solute/water\ interface}$ is larger than $\Delta G_{Water}$ (or the radius of solute is larger than Rc), this is termed as the hydrophobic solvation process,

$$\Delta G_{Solute/water\,interface} > \Delta G_{Water} \quad (R > R_c) \qquad (6)$$

In this solvation process, the hydration free energy is slightly independent from the size of dissolved solutes (Figure 3). This is different from the obvious dependence of $\Delta G_{Hydration}$ on solute size in the initial solvation process.

From equation (3), the hydration free energy is the sum of both $\Delta G_{Solute/water\,interface}$ and $\Delta G_{Water}$, and $\Delta G_{Water}$ is a constant at specific temperature and pressure. In hydrophobic solvation process, $\Delta G_{Solute/water\,interface}$ is larger than $\Delta G_{Water}$. To be more thermodynamically stable (minimizing $\Delta G_{Hydration}$), this can be achieved by minimizing $|\Delta G_{Solute/water\,interface}|$ ($\Delta G_{Solute/water\,interface}$ is negative). The $\Delta G_{Solute/water\,interface}$ is proportional to the ratio of surface area to volume of solutes (1/R), therefore the dissolved bubbles tend be accumulated to minimize the surface area (Figure 4). Of course, this undoubtedly results in hydrophobic effects. In addition, it can be derived that driving force should originate from the hydrogen bonding in bulk water rather than interfacial water.

The hydration free energy in the hydrophobic process is related to hydrogen bondings in bulk water and interfacial water. Therefore, hydrophobic effects should be closely related to the competition between hydrogen bondings in bulk water and those in interfacial water. From figure 2, in comparison with interfacial water, besides DDA, DAA and DA hydrogen bondings, DDAA (tetrahedral) hydrogen bonding can also be found in bulk water. Therefore, hydrophobic effects should be ascribed to DDAA hydrogen bondings in bulk water. In reference with interfacial water, due to the existence of DDAA hydrogen bonding in bulk water, this leads to $\Delta G_{Water}$ is lower than $\Delta G_{Solute/water\,interface}$. Based on our recent study [40], the Gibbs free energy of DDAA hydrogen bonding is mainly ascribed to high hydrogen bonding energy of the structural motif, therefore the hydrophobic process is thermodynamically described to be an enthalpic process.

From this study, the hydrophobic effects are ascribed to the competition between hydrogen bondings in bulk water ($\Delta G_{Water}$) and those in interfacial water ($\Delta G_{Solute/water\ interface}$). This means that the hydrophobic effects should result from the difference between $\Delta G_{Water}$ and $\Delta G_{Solute/water\ interface}$, which can be determined (Figure 5). Additionally, to investigate the enthalpy and entropy changes during solvation, the enthalpy difference between $\Delta H_{Water}$ and $\Delta H_{Solute/water\ interface}$, and entropy difference between $\Delta S_{Water}$ and $\Delta S_{Solute/water\ interface}$ are also calculated, respectively (Figure 5). From figure 5, it can be derived that the drive force in the initial solvation is different from that of hydrophobic solvation. For the initial solvation process, this can be ascribed to an entropic process. However, the hydrophobic process is thermodynamically described to be an enthalpic process.

## 4. Properties of hydrophobic interactions

Based on this study, hydrophobic effects should be closely related to the hydrogen bonding of liquid water, which can be ascribed to the competition between hydrogen bondings in bulk water and those in interfacial water. In the process of hydrophobic effects, the attractive force between non-polar molecules or molecular surfaces can be ascribed to DDAA hydrogen bonding of water. This is different from chemical bond, such as covalent bond or ion bond. Although the term hydrophobic bond is still used, hydrophobicity is reasonably considered as the effects rather than bond.

At equilibrium, the Gibbs free energy of hydrophobic interaction is the difference between the hydration free energies of the systems before and after the interaction. From equation (3), the

Gibbs free energy of hydrophobic effects can be determined,

$$\Delta G_{Hydrophobicity} = -\Delta G_{Air/water\,interface} \quad (7)$$

Therefore, the energy of hydrophobic interaction is closely related to the free energy of interface water.

From the above discussion on hydration free energy, with increasing radius of solute, it can be divided into initial and hydrophobic solvation processes, respectively. In the initial salvation process, the solutes are surrounded by interfacial water, and dispersed in water. It seems that there exists repulsive force between solutes. However, in the hydrophobic solvation process, the non-polar molecules or molecular surfaces are repelled to be accumulated in water. It seems that there exists attractive force between the dissolved solute. Therefore, in reference to the critical radius (Rc), the hydrophobic interaction can be divided into repulsive force (less than Rc) and attractive force (larger than Rc), which corresponds to initial and hydrophobic solvation processes, respectively (Figure 6).

Based on the equation (7), the strength of hydrophobic interaction is proportional to 1/R. If the distance is larger than Rc, there exists attractive force between dissolved solutes. Therefore, the hydrophobic interaction can be regarded as long-range attractive force. With decreasing the radius of solute, this leads to the slow increase of hydrophobic interaction. However, if the distance is less than Rc, the rapid increase of hydrophobic interaction can be observed, and can be regarded as repulsive force. In fact, this can be demonstrated by the experimentally observed force-distance relationships. In Hammer et al. study [24], hydrophobic force is divided into three regimes. The only unexpectedly strong attractive force measured in all experiments so far has a range of 100-200 Å, increasing roughly exponentially down to 10-20 Å and then more steeply down to adhesive contact at $D = 0$ or, for power-law potentials, effectively at $D \approx 2$ Å [24]. However, according to recent ultra-high resolution frequency-modulation atomic force microscope (FM-AFM) study [57], Schlesinger and Sivan measured the dependence of hydrophobic interaction on the distance between AFM tip and the surface, and discovered that the commonly observed attraction at 3-10 nanometer distances turns into pronounced repulsion below 0.3-3 nanometers. This is in agreement with our theoretical analysis on hydrophobic interactions.

When a non-polar solute is added into water, it mainly affects the topmost water layer at the interface between solute and water, and the energy of hydrophobic interaction is proportional to the ratio of surface area to volume of dissolved solute. Therefore, hydrophobic interaction should be closely related to the geometric shape of dissolved solute. If the solute is regarded as an ideal sphere, the ratio of surface to volume of the solute can be expressed as,

$$(Surface area/Volume)_{Solute} = G \cdot (Surface area/Volume)_{Sphere} \quad (8)$$

where G means geometric factor of solute, and is no less than 1. Of course, G equals to 1 for an ideal sphere. To be more thermodynamically stable, the dissolved solutes will be attracted to be aggregated in the direction where G being maximum. This indicates that the strength of hydrophobic interaction should be orientation-dependent. Of course, this is different from covalent bond. Therefore, in hydrophobic solvation process, the tendency of non-polar molecules or molecular surfaces to aggregate prefers specific direction (Figure 7). This may be applied to investigate biomolecular recognition, structure-guided ligand design, and "lock and key" metaphor.

Generally, the strength of ionic bond is often 20 kJ·mol$^{-1}$, the energy of a typical single covalent bond is 320 kJ·mol$^{-1}$, and van der waal interaction lies in the range of 0.4-4.0 kJ·mol$^{-1}$. Regarding to strength of hydrophobic interaction, this is generally less than 40 kJ·mol$^{-1}$ (Figure 6). However, the energy of hydrophobic interaction is long-distance attractive force, which is 0.4 kJ·mol$^{-1}$ at 10 nm, and 0.2 kJ·mol$^{-1}$ at 20 nm, respectively. Therefore, when hydrophobic solutes are dissolved into water, the solutes are firstly attracted by long-range hydrophobic interaction to be close to each other, and the dissolved solutes are affected by short-range interactions, such as van der waals interactions, electrostatic force and covalent bond. Therefore, hydrophobic interactions undoubtedly affect the dissolved behaviors of non-polar solutes.

## 5. Conclusions

The strength of hydrogen bonding in water is stronger than that of van der waals interaction, therefore water may play an important role in the process of hydrophobic effects. When a hydrophobic solute is dissolved into water, an interface appears between the solute and water, which may affect the hydrogen bonding of water. According to our recent structural studies on water and air/water interface, the hydrophobic solute mainly affects the hydrogen bonding of interfacial water (the topmost water layer at solute/water interface), and the loss of DDAA (tetrahedral) hydrogen bonding in interfacial water should be closely related to hydrophobic effects. From this, the hydration free energy is derived, and utilized to investigate the dissolved behaviors of hydrophobic solutes. With increasing the solute size, this leads to the increase of hydration free energy, which can be divided into the initial and hydrophobic solvation processes, respectively. In the initial solvation process, the hydration free energy is dominated by the hydrogen bonding in interfacial water. However, in the hydrophobic solvation process, the hydration free energy is related to hydrogen bondings of both bulk water and interfacial water.

Therefore, various dissolved behaviors of solutes can be expected for different solvation processes. From this study, it can be derived that hydrophobic effects originate from the competition between hydrogen bondings in bulk water and those in interfacial water.

**Acknowledgements**

This work is supported by the National Natural Science Foundation of China (Grant Nos. 41373057).

**References**

[1]   J.B. Schlenoff, A.H. Rmaile, C.B. Bucur, J. Am. Chem. Soc. 130 (2008) 13589.

[2]   A. Honciuc, D.K. Schwartz, J. Am. Chem. Soc. 131 (2009) 5973.

[3]   N. Giovambattista, P.G. Debenedetti, P.J. Rossky, Proc. Natl. Acad. Sci. U.S.A. 106 (2009) 15181.

[4]   S.M. Biros, E.C. Ullrich, F. Hof, L. Trembleau, J. Rebek, J. Am. Chem. Soc. 126 (2004) 2870.

[5]   H.S. Frank, M.W. Evans, J. Chem. Phys. 13 (1945) 507.

[6]   W. Kauzmann, Adv. Protein Chem. 14 (1959) 1.

[7]   K.A. Dill, S. Bromberg, Molecular driving forces. Garland Science, New York, 2003.

[8]   F.H. Stillinger, J. Solution Chem. 2 (1973) 141.

[9]   N.T. Southall, K.A. Dill, J. Phys. Chem. B 104 (2000) 1326.

[10]  G. Graziano, B. Lee, Biophys. Chem. 105 (2003) 241.

[11]  H.S. Ashbaugh, L.R. Pratt, Rev. Mod. Phys. 78 (2006) 159.

[12]  K. Lum, D. Chandler, J.D. Weeks, J. Phys. Chem. B 103 (1999) 4570.

[13]  D.M. Huang, D. Chandler, Proc. Natl. Acad. Sci. U.S.A. 97 (2000) 8324.

[14]  D.M. Huang, P.L. Geissler, D. Chandler, J. Phys. Chem. B 105 (2001) 6704.

[15]  P.R. ten Wolde, D. Chandler, Proc. Natl. Acad. Sci. U.S.A. 99 (2002) 6539.

[16]  D. Chandler, Nature 437 (2005) 640.

[17]  M.V. Athawale, G. Goel, T. Ghosh, T.M. Truskett, S. Garde, Proc. Natl. Acad. Sci. U.S.A. 104 (2007) 733.

[18] S. Rajamani, T.M. Truskett, S. Garde, Proc. Natl. Acad. Sci. U.S.A. 102 (2005) 9475.

[19] Y.S. Djikaev, E. Ruckenstein, J. Chem. Phys. 139 (2013) 184709.

[20] T. Lazaridis, eLS (2013) DOI: 10.1002/9780470015902.a0002974.pub2.

[21] P. Setny, R. Baron, J.A. McCammon, J. Chem. Theory Comput. 6 ( 2010) 2866.

[22] R. Baron, P. Setny, F. Paesani, J. Phys. Chem. B 116 (2012) 13774.

[23] J. Israelachvili, R. Pashley, Nature 300 (1982) 341.

[24] M.U. Hammer, T.H. Anderson, A. Chaimovich, M.S. Shell, J. Israelachvili, Faraday Discuss. 146 (2010) 299.

[25] E.E. Meyer, K.J. Rosenberg, J. Israelachvili, Proc. Natl. Acad. Sci. U.S.A. 103 (2006) 15739.

[26] S.H. Donaldson, J. Anja Røyne, K. Kristiansen, M.V. Rapp, S. Das, M.A. Gebbie, D. Woog Lee, P. Stock, M. Valtiner, J. Israelachvili, Langmuir 31 (2015) 2051.

[27] I.T.S. Li, G.C. Walker, J. Am. Chem. Soc. 132 (2010) 6530.

[28] I.T.S. Li, G.C. Walker, Proc. Natl. Acad. Sci. U.S.A. 108 (2011) 16527.

[29] I.T.S. Li, G.C. Walker, Acc. Chem. Res. 45 (2012) 2011.

[30] W.A. Ducker, D. Mastropietro, Curr. Opin. Colloid In. 22 (2016) 51.

[31] R.F. Tabor, F. Grieser, R.R. Dagastine, D.Y.C. Chan, Phys. Chem. Chem. Phys. 16 (2014) 18065.

[32] R.F. Tabor, C. Wu, F. Grieser, R.R. Dagastine, D.Y.C. Chan, J. Phys. Chem. Lett. 4 (2013) 3872.

[33] C. Shi, D.Y.C. Chan, Q. Liu, H. Zeng, J. Phys. Chem. C 118 (2014)118.

[34] H. Zeng, C. Shi, J. Huang, L. Li, G. Liu, H. Zhong, Biointerphases 116 (2016) 018903.

[35] P. Stock, T. Utzig, M. Valtiner, J. Colloid Interface Sci. 446 (2015) 244.

[36] P.E. Fraley, K.N. Rao, J. Mol. Spectrosc. 29 (1969) 348.

[37] R. Ludwig, Phys. Chem. Chem. Phys. 4 (2002) 5481.

[38] S.S. Xantheas, Chem. Phys. 258 (2000) 225.

[39] Q. Sun, Vib. Spectrosc. 51 (2009) 213.

[40] Q. Sun, Chem. Phys. Lett. 568 (2013) 90.

[41] Q. Sun, Vib. Spectrosc. 62 (2012) 110.

[42] A.K. Soper, M.A. Ricci, Phys. Rev. Lett. 84 (2000) 2881.

[43] A.M. Saitta, F. Datchi, Phys. Rev. E 67 (2003) 20201.

[44] X. Wei, Y.R. Shen, Phys. Rev. Lett. 86 (2001) 4799.

[45] A.M. Jubb, W. Hua, H.C. Allen, Acc. Chem. Res. 45 (2012) 110.

[46] Y.R. Shen, V. Ostroverkhov, Chem. Rev. 106 (2006) 1140.

[47] G.L. Richmond, Chem. Rev. 102 (2002) 2693.

[48] M. Sovago, R.K. Campen, H.J. Bakker, M. Bonn, Chem. Phys. Lett. 470 (2009) 7.

[49] M. Sovago, R.K. Campen, G. Wurpel, M. Muller, H.J. Bakker, M. Bonn, Phys. Rev. Lett. 100 (2008) 173901.

[50] C.S. Tian, Y.R. Shen, J. Am. Chem. Soc. 131 (2009) 2790.

[51] Q. Sun, Y. Guo, J. Mol. Liq. 213 (2016) 28

[52] N. Ji, V. Ostroverkhov, C.S. Tian, Y.R. Shen, Phys. Rev. Lett. 100 (2008) 096102.

[53] V. Ostroverkhov, G.A. Waychunas, Y.R. Shen, Phys. Rev. Lett. 94 (2005) 046102.

[54] C.S. Tian, Y.R. Shen, Chem. Phys. Lett. 470 (2009) 1.

[55] N.E. Dorsey, Properties of ordinary water substance, ACS Monograph No. 81, Reinhold, New York, 1940.

[56] J. Kim, Y. Tian, J.Z. Wu, J. Phys. Chem. B 119 (2015) 12108.

[57] I. Schlesinger, U. Sivan, arXiv preprint arXiv: 1603.08215 (2016).

**Fig. 1.** (a) The Raman OH stretching band of water can be deconvoluted into five sub–bands, and ascribed to OH vibrations engaged into various local hydrogen bondings. (b) The Raman OH stretching bands of water from 278 K to 328 K under 0.1 MPa. (c) For the Raman OH stretching bands of water from 278 K to 328 K, the dependence of $\ln(I_{free-OH}/I_{DDAA-OH})$ on $1/T$. The linear fit is shown in solid line.

**Fig. 2.** Structural changes across the solute/water interface. As a hydrophobic solute is dissolved into water, the solute mainly affects the structure of interfacial water (topmost water layer at the interface).

**Fig. 3.** The hydration free energy increases with the radius of dissolved solute, and can be divided into initial and hydrophobic solvation processes, respectively. At 293K and 0.1MPa, the structural transition occurs when radius of solute (critical radius, Rc) is 6.5 Å, and is shown in dashed line.

**Fig. 4.** The dissolved behaviors of solutes in the initial and hydrophobic solvation processes.

**Fig. 5.** The dependence of thermodynamic contributions on the radius of solute.

**Fig. 6.** The dependence of hydrophobic interaction on the radius of solute. In reference to the critical radius of solute (Rc), this can be divided into the short–range repulsive force and long–range attractive force, respectively.

**Fig. 7.** The effects of different interactions on dissolved behaviors of solutes. The hydrophobic interaction is proportional to the surface area to volume of solute, therefore the interaction should be orientation-dependent. Due to long-range attractive hydrophobic interaction, the solutes approach to be close to each other, the dissolved behaviors are modulated by van der waals interactions, electrostatic force and covalent bond.

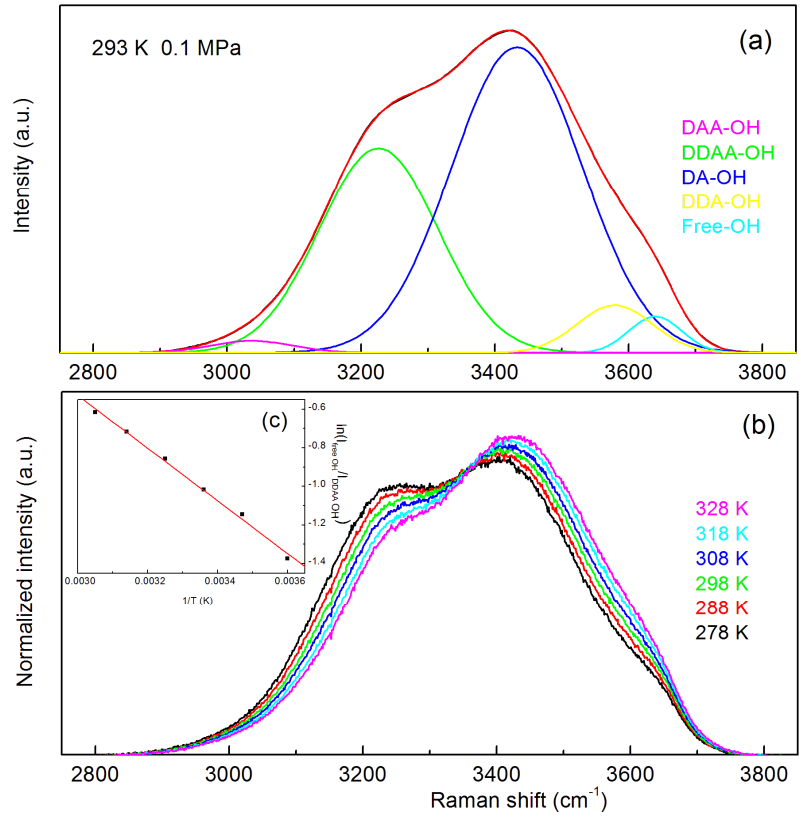

Fig. 1.

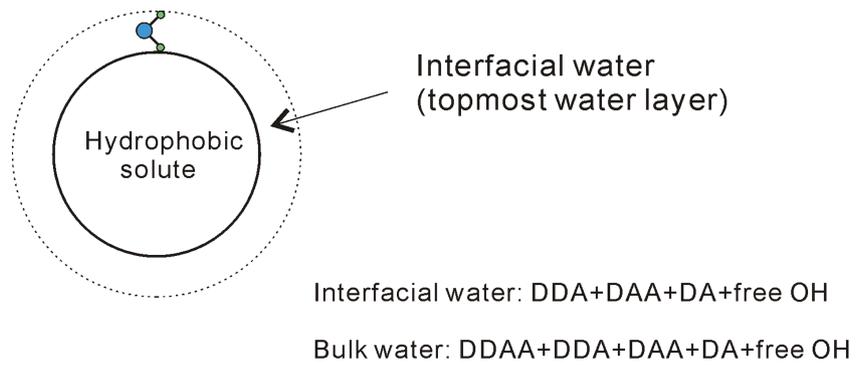

Interfacial water: DDA+DAA+DA+free OH

Bulk water: DDAA+DDA+DAA+DA+free OH

Loss of DDAA (tetrahedral) HB in interfacial water

Gibbs free energy of solute/water interface

Fig. 2.

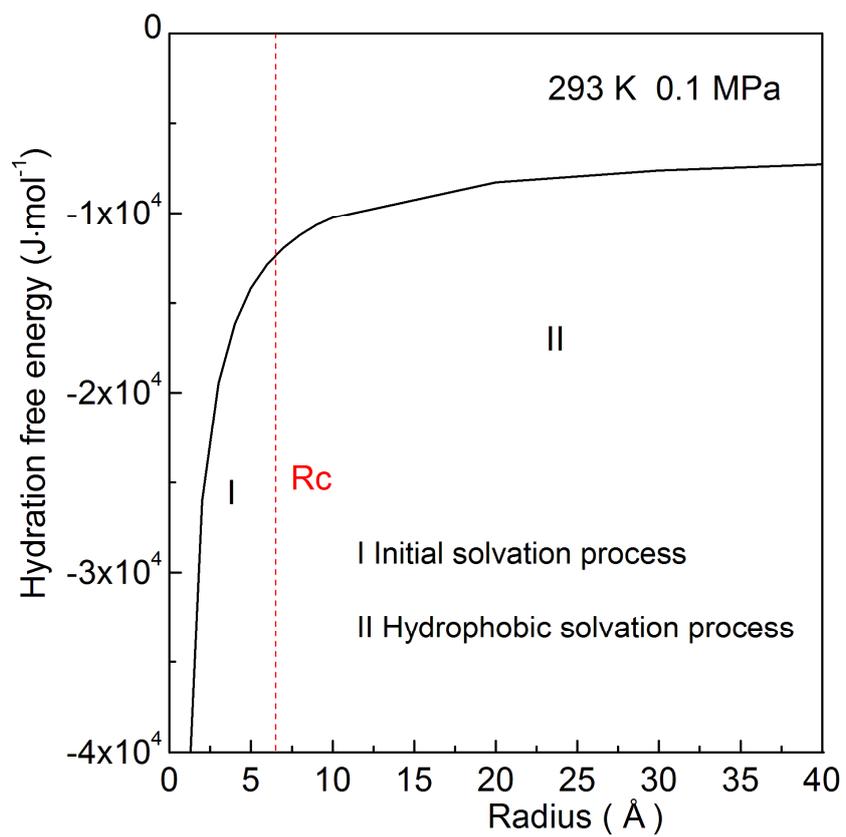

Fig. 3.

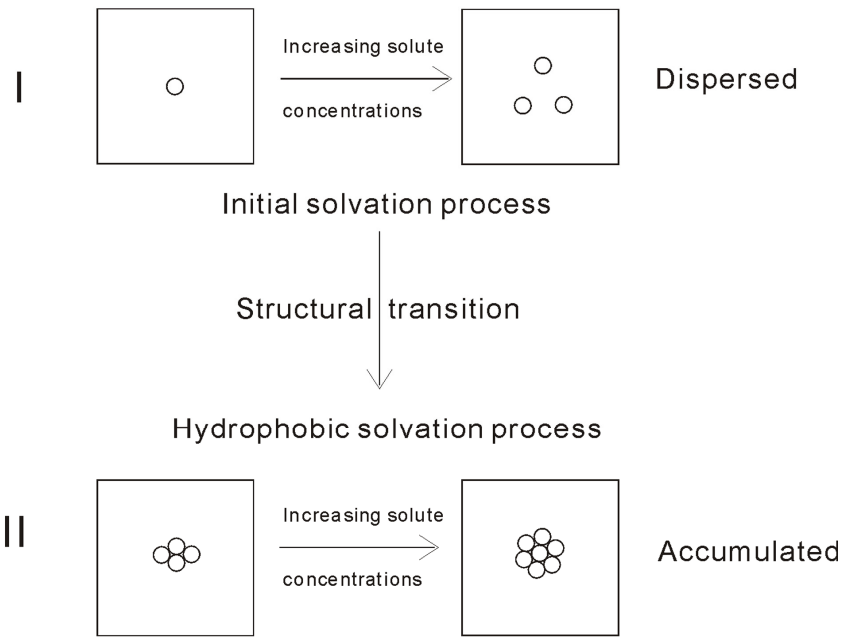

Fig. 4.

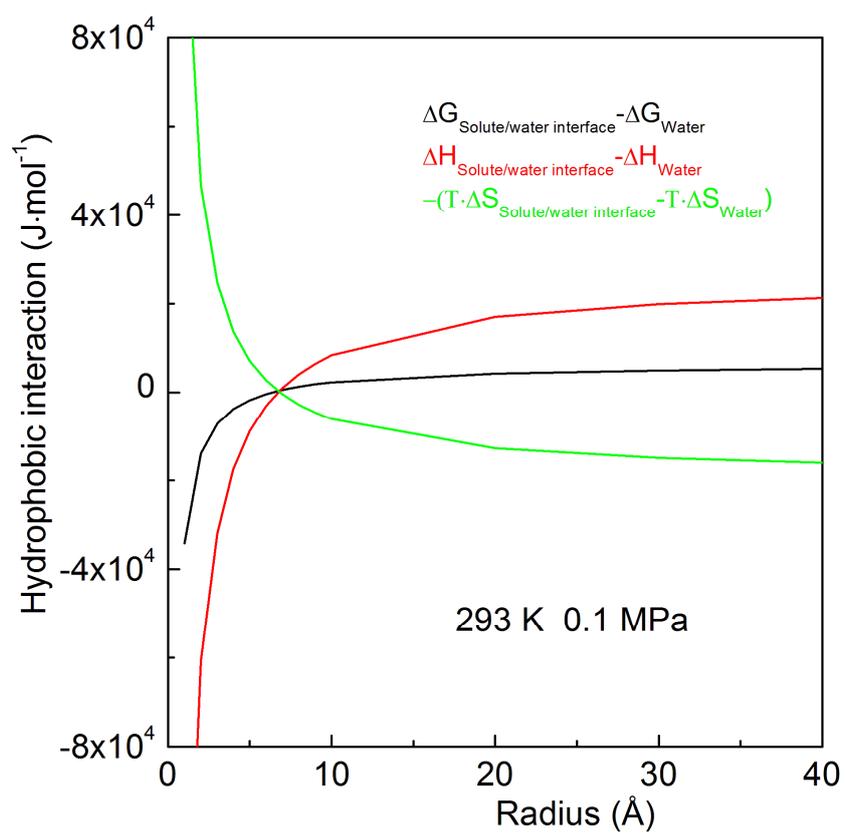

Fig. 5.

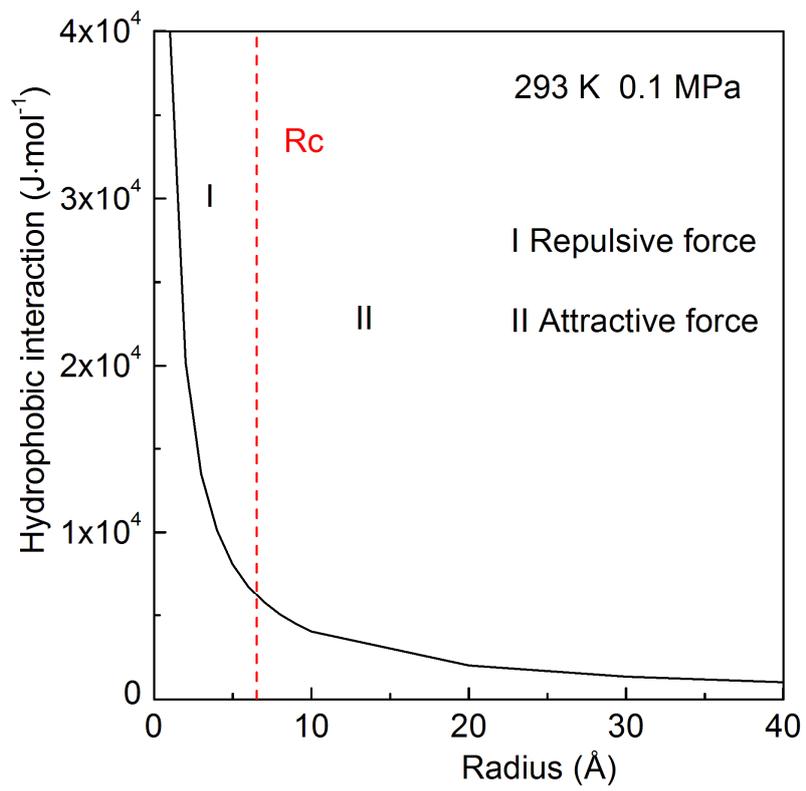

Fig. 6.

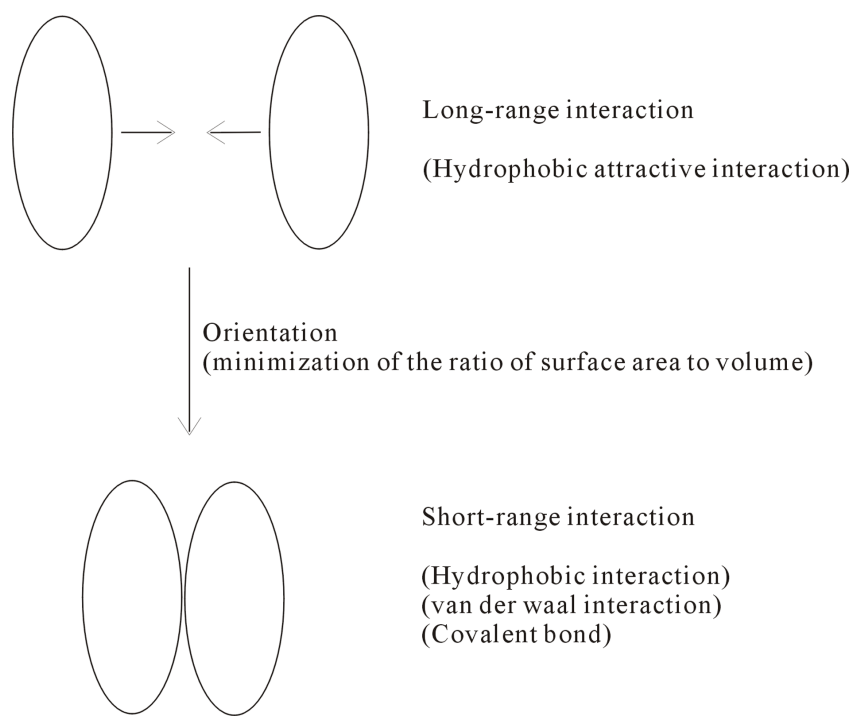

Fig. 7.